\documentclass[preprint,nofootinbib,11pt,floatfix,superscriptaddress,aps,prd,showpacs]{revtex4-2}
\usepackage{graphicx}
\usepackage{dcolumn}
\usepackage{bm}
\usepackage{hyperref}
\usepackage{color,xcolor}
\usepackage{bigints}
\usepackage{mathtools} 

\allowdisplaybreaks

\begin{document}

\title{Traversable double-throat wormholes in a string cloud background}

\author{Yvens Amaral} \email{yvens.silva@aluno.uece.br}\author{M. S. Cunha}\email{marcony.cunha@uece.br}
\affiliation{Universidade Estadual do Ceará (UECE), Centro de Ciências e Tecnologia (CCT), CEP 60714-903, Fortaleza-CE, Brazil}
\author{C.R. Muniz}\email{celio.muniz@uece.br}\affiliation{Universidade Estadual do Ceará (UECE), Faculdade de Educação, Ciências e Letras de Iguatu (FECLI), CEP 63502-253, Iguatu-CE, Brazil.} \author{M. O. Tahim}\email{makarius.tahim@uece.br}\affiliation{Universidade Estadual do Ceará (UECE), Faculdade de Educação, Ciências e Letras do Sertão Central (FECLESC), CEP 63900-000, Quixadá-CE, Brasil.}

\date{\today}

\begin{abstract}
This work constructs a new class of traversable wormhole solutions with a double-throat topology, modeled as a localized perturbation of the Ellis-Bronnikov metric in a string cloud background. Embedding diagrams and the analysis of curvature invariants, including the Kretschmann scalar and the Weyl invariant, illustrate the geometric transition from single to double-throat structures as a function of the perturbation amplitude. By imposing the zero-tidal condition, we derive analytical expressions for the energy density and pressures, showing an asymptotic $r^{-2}$ decay characteristic of a string cloud, endowed with the topology of a global monopole. A key finding is that the energy density converges to a positive constant at the center, with the radial pressure becoming negative. This local behavior provides the repulsive support necessary to inflate the interthroat region with non-exotic matter, concentrating Null Energy Condition violations to the throat vicinities. These results suggest that multi-throat geometries offer a natural mechanism for localizing exotic matter while maintaining a physical asymptotic background.
\end{abstract}

\maketitle

\newpage

\section{Introduction}

Traversable wormholes are hypothetical topological structures of spacetime that connect two distinct asymptotically flat regions or two distant locations within the same universe. Since the seminal work of Morris and Thorne \cite{MorrisThorne1988}, it has been well established that the maintenance of a wormhole throat against gravitational collapse requires the violation of the Null Energy Condition (NEC) in the framework of General Relativity \cite{Visser1995}. Matter sources exhibiting such behavior are generically referred to as exotic matter, raising fundamental questions about the physical plausibility of these geometries.

As a consequence, a central theme in wormhole physics has been the search for mechanisms capable of minimizing, localizing, or geometrically confining NEC violations. The fundamental geometrical requirements for traversable wormholes, including the flare-out condition at the throat and its intimate relation with energy condition violations, were systematically clarified in the seminal works of Visser and collaborators \cite{Visser1995, HochbergVisser1998}. Comprehensive reviews of wormhole physics within General Relativity and beyond can be found in Ref.~\cite{Lobo2005Review}. 

Several approaches have been proposed in the literature, including thin-shell constructions \cite{Lobo2025,ThinShellWHdynamic_Lobo_Visser_Garcia,ThinShellSWH_Xiaobiao}, matter sources \cite{munizWH2026,Axion_boson_star_wormhole_ChenHao}, and extensions of General Relativity \cite{hassan2025, Radhakrishnan2024,Antoniou2020,HOCHBERG1990349,VictorioFalco_Extendedgravity,VictorioFalco_Wormholes_metricaffine}. Among exact solutions within standard gravity, the Ellis–Bronnikov wormhole \cite{Ellis1973,Bronnikov1973} stands as the paradigmatic example of a traversable wormhole supported by a phantom scalar field. Owing to its simplicity and zero–tidal–force nature, this geometry has served as a valuable theoretical laboratory for investigating geodesics, lensing, stability, and quantum effects \cite{SWH_Takayuki_Noboyuki,ThinShellSWH_Xiaobiao,Double_Shadow_WH_Maciek}.

More recently, generalized versions of the Ellis–Bronnikov solution have been introduced, commonly referred to as Generalized Ellis–Bronnikov (GEB) wormholes \cite{Kar:1995jz,GEB_nonlineareletro_XinSu}. These models incorporate an additional geometric parameter that controls the steepness of the throat and the curvature distribution near the core, enabling detailed studies of optical phenomena such as gravitational lensing and shadow formation \cite{Crispim2025,Sokoliuk2022}. Importantly, these generalizations allow for nontrivial internal geometries without spoiling asymptotic flatness. 

Parallel to these developments, it has been recognized that wormhole geometries may admit even more complex internal structures. In particular, several studies have shown that single-throat configurations can transition into double-throat structures featuring an intermediate equatorial or belly region. Such configurations have been reported in wormholes threaded by chiral fields \cite{Kunz2013Chiral}, in Einstein–vector–Gauss–Bonnet theory \cite{Kunz2022EVGB}, in the framework of General Relativity via construction \cite{Cataldo_doublethroat}, in Einstein–scalar–Gauss–Bonnet gravity without exotic matter \cite{Doneva2020ESGB}, in systems with axion and phantom fields in nonlinear electrodynamics \cite{Axion_boson_star_wormhole_ChenHao,Crispim_2025} and in non-Abelian wormholes supported by Yang–Mills–Higgs fields \cite{Kunz2020NonAbelian}. In these scenarios, the emergence of a double-throat is often related with strong coupling effects, higher-curvature terms, or nontrivial matter interactions, suggesting that multi-throat geometries are a generic feature of gravitating systems rather than isolated curiosities.

Despite this progress, the physical interpretation of the matter sources supporting wormholes remains an open problem. A particularly appealing candidate is the string cloud model, originally proposed by Letelier \cite{Letelier1979}, which describes a fluid of radially oriented, disordered cosmic strings. Such configurations naturally generate spacetimes with a global deficit angle and an energy density decaying as $r^{-2}$, a behavior commonly encountered in wormhole solutions. String cloud backgrounds have also been extensively discussed in connection with global monopoles and topological defects \cite{BarriolaVilenkin1989}.

In this work, we propose a novel wormhole model that unifies these ideas within standard General Relativity. We construct a traversable wormhole with a nontrivial double-throat topology embedded in an Ellis–Bronnikov geometry that asymptotically matches a string cloud background. The zero–tidal–force condition is imposed in order to guarantee safe traversability and analytical control of the field equations. Within this framework, the exotic matter required to sustain the throats is confined to compact regions near each minimal surface, while the spacetime asymptotically approaches physically a string cloud configuration.

A key result of this study is that the transition to a double-throat wormhole is not the result of fine tuning. Instead, we show that for generalized geometries the formation of two throats arises naturally from localized deformations of the core. Moreover, by allowing the interthroat region to be arbitrarily large and supported by non-exotic matter (NEC-respecting), the model effectively minimizes the total volume of NEC-violating matter. This behavior stands in sharp contrast to single-throat wormholes, where exotic matter typically extends over a larger region of spacetime.

These results indicate that multi-throat wormholes can emerge as energetically favored configurations, providing a geometrically natural mechanism for the localization of exotic matter and opening new possibilities for the study of wormhole optics, particle dynamics, and stability within General Relativity.

The paper is organized as follows. In Sec.~II we introduce the spacetime geometry and present the Ellis--Bronnikov framework under the zero--tidal--force condition. In Sec.~III we analyze the emergence of the double--throat structure and discuss the geometrical conditions leading to this topology. The matter content, its physical interpretation in terms of a string cloud background, the relationship between the density in the core, the belly formation and the concentration of NEC violations in regions near the throats are discussed in Sec.~III. Finally, the last section is devoted to our concluding remarks.

\section{Metric, Field Equations, and Model Geometry}\label{field_eqs}

This section establishes the mathematical framework for the proposed double-throat wormhole. We introduce the static, spherically symmetric metric in proper radial distance coordinates and derive the corresponding Einstein field equations for an anisotropic fluid associated with the Letelier cloud string, codified in a deficit factor of the solid angle. Furthermore, we specify the model geometry through a perturbed Ellis-Bronnikov (EB) shape function, analyzing the conditions for single and double-throat formation.

\subsection{Metric and Field Equations}

We consider a static and spherically symmetric spacetime described by the line element written in terms of the proper radial distance $\ell$,
\begin{equation}
    ds^2 = -e^{2\Phi(\ell)} dt^2 + d\ell^2 + r(\ell)^2 d\Omega^2,
    \label{eq:metric}
\end{equation}
where $\ell \in (-\infty,+\infty)$, $r(\ell)$ is the shape function encoding the spatial geometry, and $\Phi(\ell)$ is the redshift function. The use of the proper radial distance provides a global and regular description of the wormhole geometry, particularly suitable for configurations admitting multiple throats.

In this coordinate system, a wormhole throat corresponds to a local minimum of the areal radius, satisfying $r'(\ell)=0$ and $r''(\ell)>0$. More general configurations may admit multiple such minima, separated by intermediate regions where $r(\ell)$ reaches a local maximum, giving rise to a double--throat structure.

To ensure that observers traversing the wormhole experience no tidal forces, we impose the zero--tidal--force condition $\Phi(\ell)=0$, which implies a constant redshift function and $g_{tt}=-1$ throughout the spacetime. This assumption significantly simplifies the field equations and allows for a transparent geometrical interpretation of the energy conditions. We emphasize that the analysis presented here is restricted to this class of wormhole geometries; nevertheless, we expect the qualitative features associated with the concentration of null energy condition violations near the throats to remain robust under small deviations from a constant redshift function.

The matter content is modeled as an anisotropic fluid described by the stress--energy tensor
\begin{equation}
T^\mu_{\ \nu} = \mathrm{diag}(-\rho,\, p_r,\, p_t,\, p_t),
\end{equation}
where $\rho$ is the energy density, $p_r$ the radial pressure, and $p_t$ the tangential pressure. Under these assumptions, Einstein’s field equations $G_{\mu\nu}=8\pi T_{\mu\nu}$ yield
\begin{align}
    8\pi \rho(\ell) &= \frac{1 - r'(\ell)^2 - 2 r(\ell) r''(\ell)}{r^2(\ell)}, \label{eq:rho_general}\\
    8\pi p_r(\ell) &= \frac{r'(\ell)^2 - 1}{r^2(\ell)}, \label{eq:pr_general}\\
    8\pi p_t(\ell) &= \frac{r''(\ell)}{r(\ell)}, \label{eq:pt_general}
\end{align}
where a prime denotes differentiation with respect to $\ell$.

In particular, the combination $\rho+p_r$, which governs the null energy condition, is directly controlled by the second derivative of the shape function. This highlights the purely geometric origin of NEC violation in zero--tidal--force wormholes and anticipates the key role played by the curvature of $r(\ell)$ in the emergence of single-- and double--throat configurations. In the next section, we specify a class of shape functions that naturally leads to a double--throat geometry embedded in a string cloud background.

\subsection{The Double-Throat Model Geometry}

We propose a shape function $r(\ell)$ constructed as a perturbation of the Ellis-Bronnikov metric, scaled by a factor $\alpha$ associated with the string cloud deficit angle. The master function is defined as:
\begin{equation}
    r(\ell) = \alpha \sqrt{\ell^2 + r_0^2}+ \dfrac{A}{1 + {\ell^2}/{b^2}} .
    \label{eq:master_function}
\end{equation}
Here, the parameters have distinct physical roles:
\begin{itemize}
       \item $\alpha \in (0,1)$: The deficit angle related to the string cloud parameter. $\alpha$ was defined based in \cite{Letelier1979} as $\alpha^2 = 1-a$
    \item $r_0$: The throat radius of the background geometry when $A=0$.
    \item $A$ and $b$: The amplitude and width of the localized perturbation responsible for the double-throat topology.
\end{itemize}

\begin{figure}[h!]
    \centering
           \includegraphics[width=0.55\linewidth]{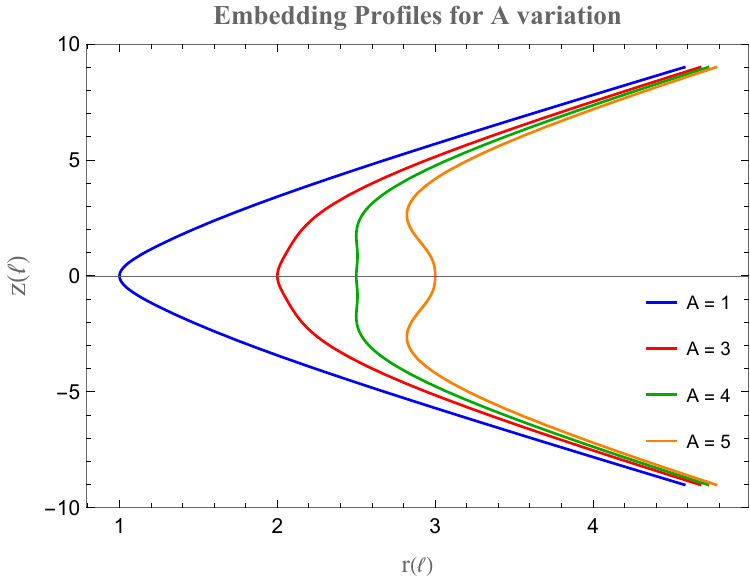}
            \includegraphics[width=0.42\linewidth]{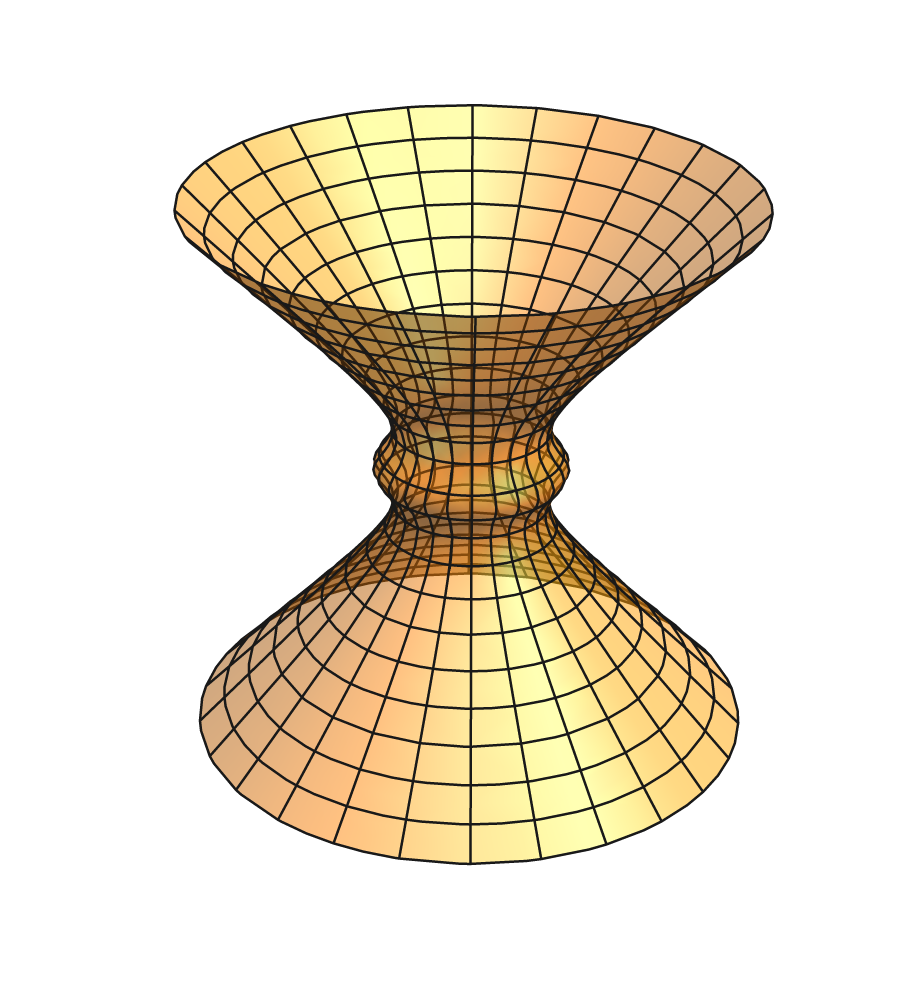}
       \caption{Geometric representation of the Ellis-Bronnikov wormhole under the influence of a string cloud and the double-throat perturbation. (Left) The 2D embedding profiles illustrate the transition from a single-throat to a double-throat geometry as the amplitude parameter $A$ increases. (Right) Corresponding 3D embedding diagram for the double-throat case ($A=5$), highlighting the formation of the central belly region situated between the two minimal throat radii. In both plots, the parameters are set to $\alpha=0.5$ and $r_0=1.0$.}.\label{fig:embedding}
    \label{rho}
\end{figure}
The emergence of a double-throat structure is characterized by the transition of the central point, $\ell=0$, from a global minimum (representing a single throat) to a local maximum (representing the interthroat region), as clearly seen in the embedding profiles of Fig. \ref{fig:embedding}. Mathematically, this bifurcation occurs when the concavity of the shape function at the origin becomes negative, \textit{i.e.}, $r''(0) < 0$.

Starting from the modified shape function defined in Eq.~(\ref{eq:master_function}), we compute the second derivative with respect to the proper radial distance $\ell$ evaluated at the origin,
\begin{equation}
r''(0) = \left[ \alpha r_0^2 (\ell^2 + r_0^2)^{-3/2} - \frac{2Ab^2(b^2 - 3\ell^2)}{(b^2 + \ell^2)^3} \right]_{\ell=0}.
\end{equation}
Thus, the expression for the second derivative of the areal coordinate $r(\ell)$ at the origin is given by:
\begin{equation}
    r''(0) = \frac{\alpha}{r_0} - \frac{2A}{b^2}.
    \label{eq:curvature_origin}
\end{equation}
The condition for the formation of a double-throat ($r''(0) < 0$) therefore leads to the regime in which the perturbation amplitude $A$ must overcome the natural tendency of the throat to close. The condition requires a threshold value:
    \begin{equation}\label{conditionA)}
        A > \frac{\alpha b^2}{2r_0} \,.
    \end{equation}

In order to visualize the geometric structure of the double-throat wormhole, specifically the phenomenology of the two throats and the central belly, we construct an embedding diagram where we can see the formation of the interthroat belly (Fig. \ref{fig:embedding}). We consider a static slice of the spacetime ($t = \text{const}$) restricted to the equatorial plane ($\theta = \pi/2$). The metric \eqref{eq:metric} reduces to the two-dimensional line element, that is,
\begin{equation}
    ds^2_{\Sigma} = d\ell^2 + r^2(\ell) d\phi^2.
\end{equation}
We embed this surface into a three-dimensional Euclidean space with cylindrical coordinates $(z, r, \phi)$, characterized by the metric $ds^2_{E} = dz^2 + dr^2 + r^2 d\phi^2$. By identifying the metrics $ds^2_{\Sigma} = ds^2_{E}$, we obtain the fundamental relation $d\ell^2 = dz^2 + dr^2$. This leads to the differential equation governing the vertical embedding coordinate $z(\ell)$:
\begin{equation}
    \frac{dz}{d\ell} = \pm \sqrt{1 - \left(\frac{dr}{d\ell}\right)^2}.
    \label{eq:embedding}
\end{equation}
The numerical integration of Eq. \eqref{eq:embedding} generates the profile curve $z(\ell)$, which, upon rotation around the $z$-axis, reveals the global topology of the wormhole. The existence of the embedding requires the condition $|dr/d\ell| \le 1$ to be satisfied throughout the domain. 

Figure \ref{fig:embedding} illustrates the geometric structure of the wormhole through two complementary perspectives. The left panel displays the 2D embedding profiles, which reveal the transition from a single-throat to a double-throat configuration as the perturbation amplitude $A$ increases. Specifically, for amplitudes above the critical threshold, the central region at $\ell=0$ transforms from a global minimum into a local maximum, giving rise to the interthroat belly. This behavior is further clarified in the right panel, which depicts the 3D surface of revolution for a representative double-throat case ($A=5$), showing the final spatial geometry resulting from rotation around the vertical symmetry axis.

A distinctive feature of the proposed model is the ability to continuously tune the geometry of the central region, transitioning from a single-throat configuration to a double-throat system without altering the asymptotic properties of the spacetime and the throat topology. To quantify this geometric response, we analyze two distinct measures of separation between the minimal surfaces located at $\ell = \pm \ell_{th}$.

First, the physical proper distance $\Delta L$ is determined by the path integral along the radial geodesic. Second, the embedding height $\Delta z$ is obtained by isometrically embedding the equatorial slice ($t=\text{const}, \theta=\pi/2$) into a three-dimensional Euclidean space with cylindrical coordinates $(z, r, \phi)$. The condition $ds^2_{induced} = dz^2 + dr^2 + r^2 d\phi^2$ leads to the differential relation $dz/d\ell = \pm \sqrt{1 - [r'(\ell)]^2}$. Consequently, the two measures are given by
\begin{equation}
    \Delta L = \int_{-\ell_{th}}^{+\ell_{th}} d\ell = 2 \ell_{th},
\end{equation}
and
\begin{equation}
    \Delta z = \int_{-\ell_{th}}^{+\ell_{th}} \sqrt{1 - \left(\frac{dr}{d\ell}\right)^2} \, d\ell = 2 \int_{0}^{\ell_{th}} \sqrt{1 - [r'(\ell)]^2} \, d\ell,
\end{equation}
where $\ell_{th}$ is the root of $r'(\ell_{th}) = 0$ in the domain $\ell > 0$. While $\Delta L$ represents the actual travel distance required for an observer to traverse the interthroat region, $\Delta z$ corresponds to the vertical separation in the embedding diagram.

\begin{figure}[h!]
    \centering
    \includegraphics[width=0.65\linewidth]{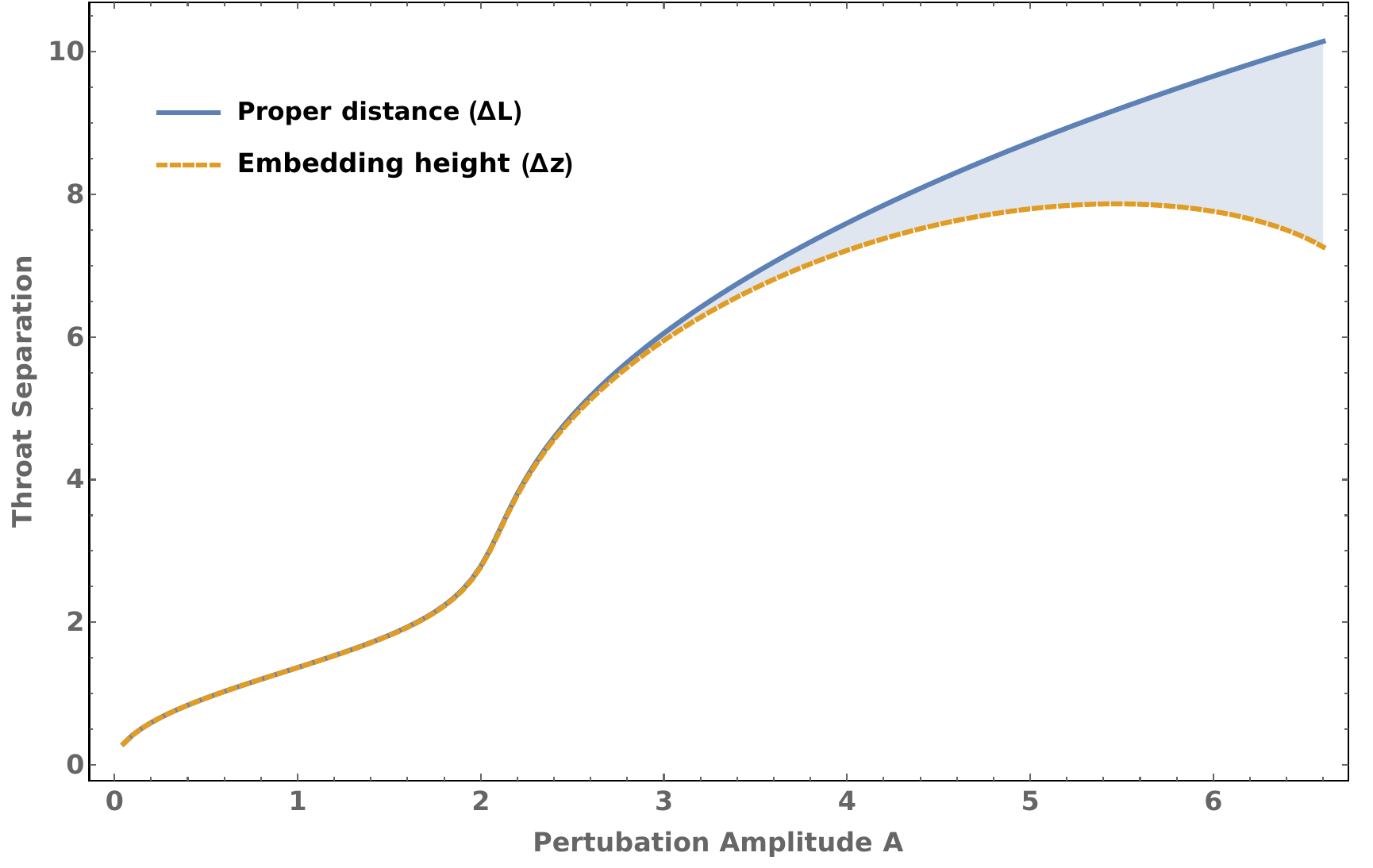} 
         \caption{Throat separation analysis as a function of the perturbation amplitude $A$ for fixed $\alpha=0.5$. The solid blue line represents the physical proper distance $\Delta L$, while the dashed orange line depicts the extrinsic embedding height $\Delta z$. Parameters are set to $r_0=1$ and $b=3$.}
    \label{fig:Separation}
\end{figure}

In Fig.~\ref{fig:Separation}, we analyze the separation between the two minimal surfaces induced by the double-throat configuration. As the perturbation amplitude $A$ increases, it is observed that the proper distance $\Delta L$ is systematically larger than the embedding height $\Delta z$. This discrepancy ($\Delta L > \Delta z$) provides a quantitative measure of the non-Euclidean curvature of the interthroat region (the \textit{belly}), indicating that the gravitational well is physically deeper than suggested by its Euclidean embedding visual representation. 

\subsection{Curvature analysis}

In order to characterize the gravitational field of the double-throat configuration we study the behavior of several curvature indicators. The Ricci's scalar curvature for the space geometry is given by
\begin{equation}
    R = 2\left[\frac{1-(r(\ell)')^2 - 2r(\ell)r(\ell)''}{r(\ell)^2}\right].
\end{equation}
The Kretschmann's scalar, $K = R_{\mu\nu\rho\sigma}R^{\mu\nu\rho\sigma}$, has the following expression
\begin{equation}
    K=4\left[\frac{2 r(\ell)^2 r''(\ell)^2 +  \left( r'(\ell)^2 - 1 \right)^2}{r(\ell)^4}\right]
\end{equation}
 and the Ricci tensor squared is
 \begin{equation}
     R_{\mu\nu}R^{\mu\nu} = \frac{4 r''(\ell)^2}{r^2(\ell)} + 2\left(\frac{  r(\ell) r''(\ell) + r'(\ell)^2 - 1 }{r(\ell)^2}\right)^2
 \end{equation}
\begin{figure}[h]
    \centering
    \includegraphics[width=0.45\linewidth]{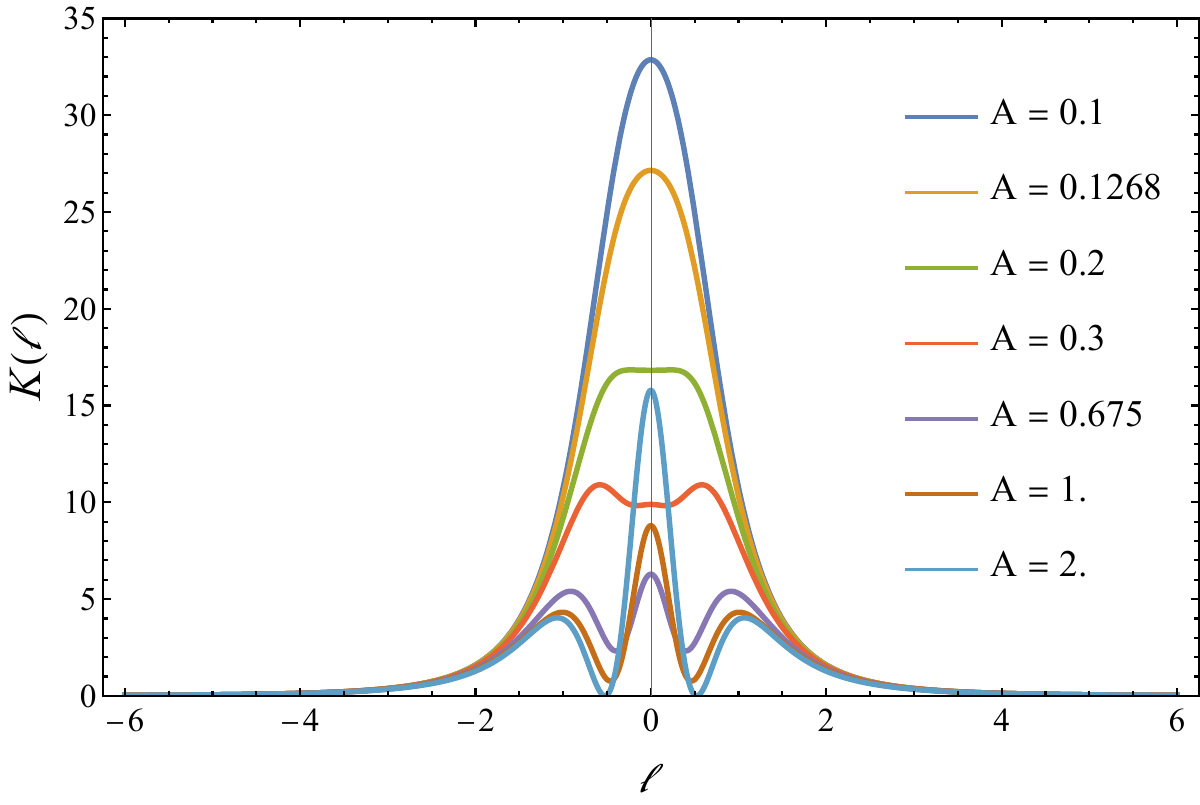}
    \includegraphics[width=0.45\linewidth]{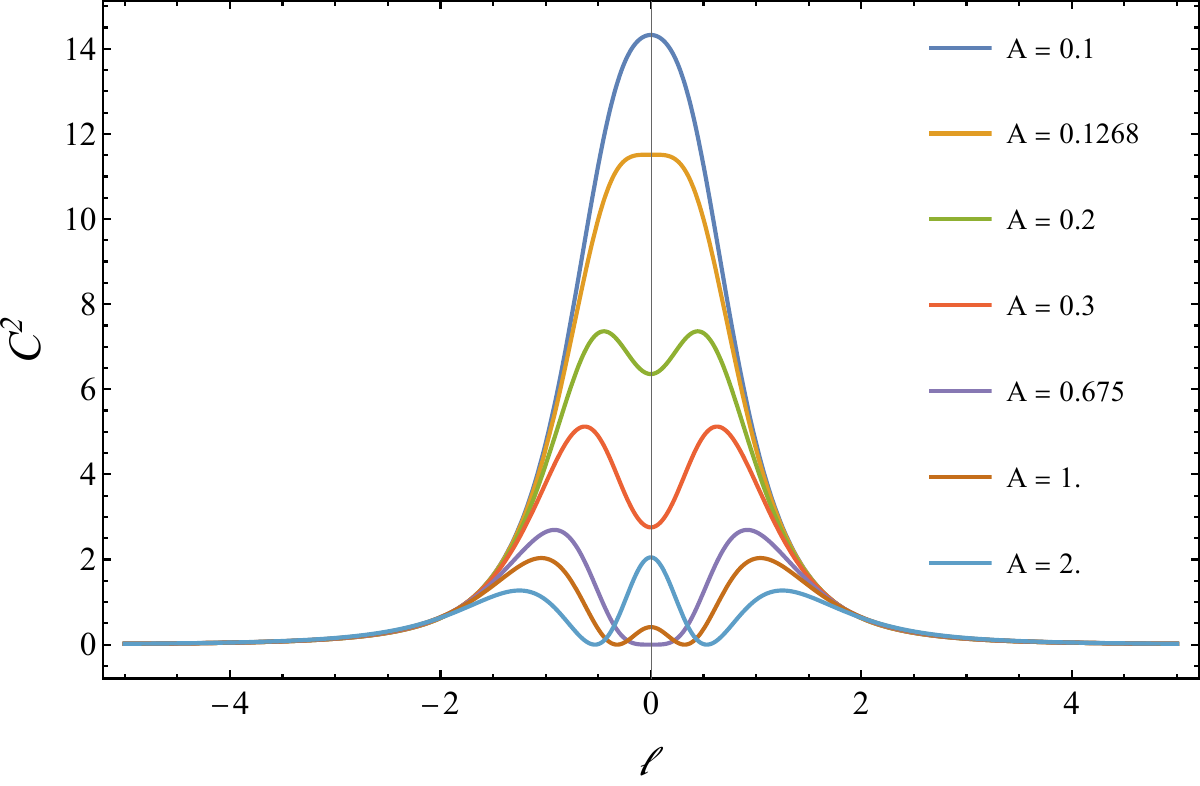}
    \includegraphics[width=0.45\linewidth]{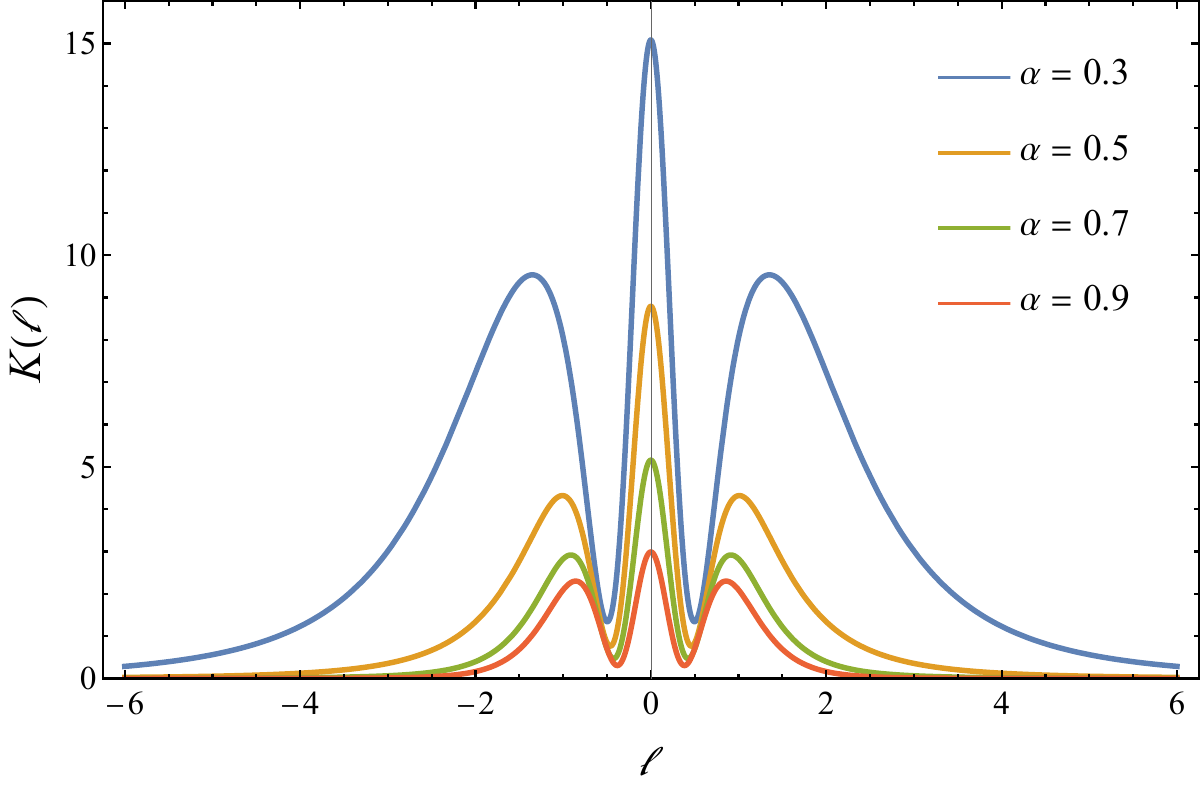}
    \includegraphics[width=0.45\linewidth]{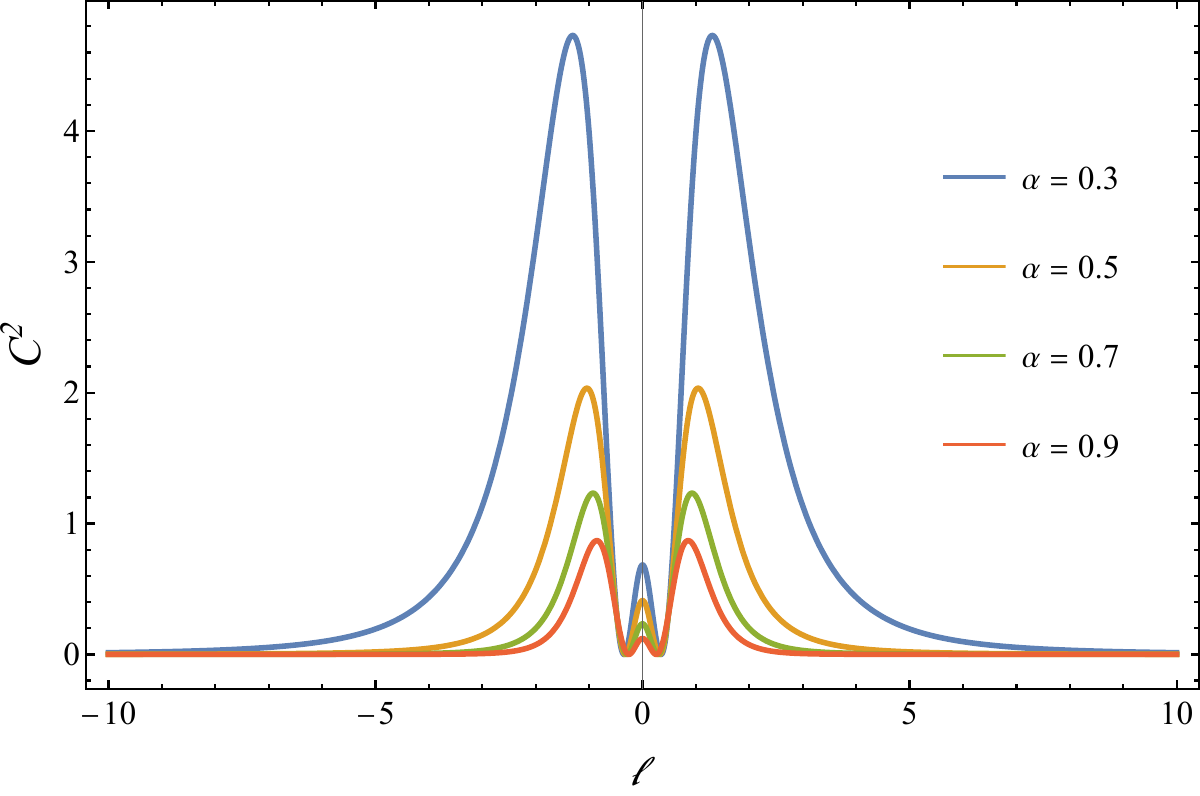}
 \caption{Comparative analysis of curvature scalars. The top panels display the Kretschmann scalar $K$ (left) and the Weyl invariant $C^2$ (right) as functions of the proper distance $\ell$ for different perturbation amplitudes $A$, for $r_0 = 1, b=1,\alpha=0.5$. The bottom panels show the behavior of $K$ (left) and $C^2$ (right) for various string cloud parameters $\alpha$ with a fixed amplitude, $A=1$. In the double-throat regime, both scalars exhibit a characteristic multi-peak signature, highlighting the intense geometric distortion at the throats and the central belly.}
    \label{fig: CURVATURE_SCALARS}
\end{figure}

In particular, we introduce the Weyl invariant, $C^2 = C_{\mu\nu\rho\sigma}C^{\mu\nu\rho\sigma}$. While the Kretschmann scalar $K$ provides a measure of the total spacetime curvature, including contributions from the local energy-momentum distribution via the Ricci tensor, the Weyl invariant isolates the purely geometric distortion or ``shape'' of the manifold. For the class of static, spherically symmetric metrics with the zero-tidal condition ($g_{tt}=-1$), the Weyl invariant is given by 
\begin{equation}
    C^2 = \frac{4}{3} \left[ \frac{r(\ell)r''(\ell)  -r'(\ell)^2 + 1}{r(\ell)^2} \right]^2,
\end{equation}
in the coordinates of proper distance. This scalar is particularly useful for analyzing the spatial tidal forces and the structural stretching required to sustain the double-throat geometry. 

The analysis of the curvature scalars, presented in Fig. \ref{fig: CURVATURE_SCALARS}, characterizes the geometric features of the model. As shown in the top panels, increasing the amplitude $A$ makes the curvature distribution more pronounced, presenting a tendency to increase especially in the belly , leading to the formation of the interthroat belly. In the limit $A \rightarrow 0$, the system recovers the standard single-throat Ellis-Bronnikov geometry and the belly region vanishes. The consistent behavior of both the Kretschmann scalar $K$ and the Weyl invariant $C^2$ confirms that this bifurcation is a structural geometric feature of the manifold. As \(A \rightarrow\infty\) the Weyl's scalar at belly's center tends to \(16/3b^4\), while the Kretschmann's scalar tends to \(32/b^2\).

Furthermore, the bottom panels illustrate a fundamental relationship between the string cloud parameter $\alpha$ and the curvature intensity. Decreasing $\alpha$ leads to a sharp increase in the curvature at the throats. This indicates that the string cloud background acts as a global confining agent, modulating the energetic scale and compactness of the wormhole core. While $A$ triggers the topological split, $\alpha$ effectively controls the curvature intensity throughout the entire double-throat structure.

\section{Matter Source and Energy Conditions}
\label{sec:energy_conditions}

This section derives the analytical expressions for the energy density and pressures of the matter source. We demonstrate that the spacetime asymptotically approaches a string cloud background and that the energy density in the core is constant, with the radial pressure becoming negative, which inflates the belly and needs to be large enough to can do it. We also investigate the Null Energy Condition (NEC) to show the concentration of exotic matter near the throats.

\subsection{Analysis of the Matter Source}

Substituting Eqs. (\ref{eq:master_function}) into the field equations (\ref{eq:rho_general}-\ref{eq:pt_general}), we derive the complete analytical expressions for the matter source components. The energy density $\rho(\ell)$ is the most complex component, incorporating contributions from both the background geometry and the local perturbation. The full expression is
\begin{align}\label{eq:rho_full}
\rho(\ell)&=-\frac{1}{8\pi\left(\dfrac{A b^{2}}{b^{2}+\ell^{2}}+\alpha\left(\ell^{2}+r_{0}^{2}\right)^{1/2}
\right)^{\!\!2}
}
\Bigg[-1 + \left(\frac{\ell}{\alpha~\left(\ell^{2}+r_{0}^{2}\right)^{1/2}}-\frac{2 A b^{2}\,\ell}{(b^{2}+\ell^{2})^{2}}\right)^{\!\!2}
\nonumber\\[0.3cm]
&+\,2\left(\frac{\alpha~r_{0}^{2}}{\left(\ell^{2}+r_{0}^{2}\right)^{3/2}}-\frac{2Ab^2(b^{2}-3 \ell^{2})}{(b^{2}+\ell^{2})^{3}}\right)
    \left(\frac{A b^{2}}{b^{2}+\ell^{2}}+ \alpha\left(\ell^{2}+r_{0}^{2}\right)^{1/2}\right)
\Bigg].
\end{align}
Let us analyze the behavior of the derived expressions in the asymptotic limit $\ell \to \infty$. In this regime, the perturbation terms decay rapidly ($\mathcal{O}(\ell^{-2})$) and the GEB term is dominated by $\ell^m$. Thus, we have
\begin{equation}
    r(\ell) \approx \alpha~\ell, \quad r'(\ell) \approx \alpha, \quad r''(\ell) \approx 0.
\end{equation}
Substituting these into Eq. (\ref{eq:rho_full}), the energy density simplifies to:
\begin{equation}
    \rho(\ell \to \infty) \approx \frac{1 - \alpha^2}{8\pi \alpha^2 \ell^2}.\label{densitycs}
\end{equation}
This $\ell^{-2}\approx r^{-2}$ decay is the characteristic signature of a string cloud (or a global monopole): when compared with the density in \cite{Letelier1979} we conclude it has the exactly same density in asymptotic regime. The constant $\alpha < 1$ ensures that the energy density remains positive at large distances.

\begin{figure}[h!]
    \centering
            \includegraphics[width=0.45\linewidth]{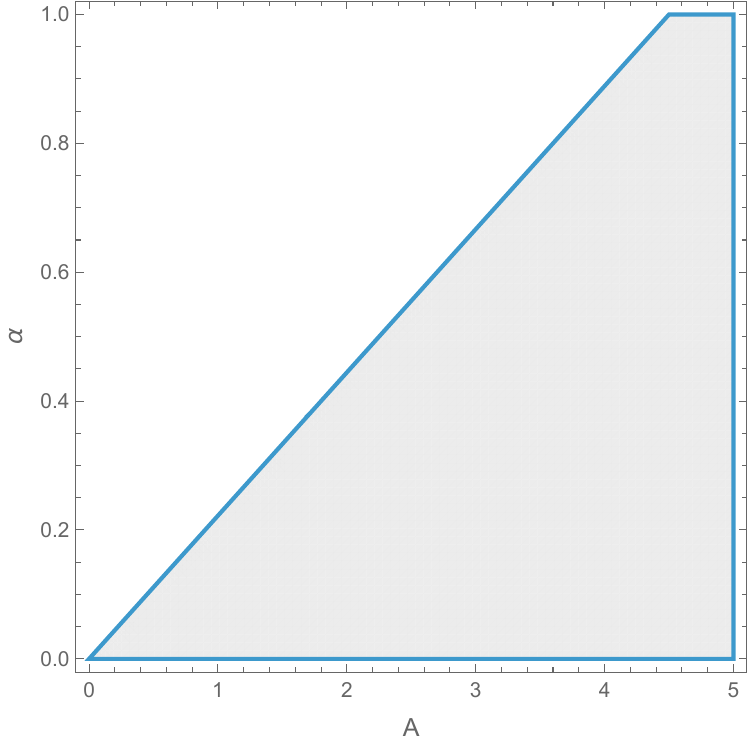}
      \caption{
Parameter space in the $(A,\alpha)$ plane for which the central energy density is positive, for $b=3.0$ and $r_0=1.0$, leading to a repulsive contribution that favors the formation of the belly. The hatched region indicates the allowed parameter values satisfying $\rho(0)>0$ and the constraint given by Eq.~(\ref{conditionA)}).
}
    \label{parameterspace2}
\end{figure}

In the opposite limit corresponding to the central region between the throats, the energy density reduces to
\begin{equation}\label{rhocentral}
\rho(\ell\to 0)\approx
\frac{
1
-
2\left(
-\frac{2A}{b^{2}}
+
\frac{\alpha}{r_{0}}
\right)
\left(
A + r_{0}\alpha
\right)
}{
8\pi\left(A + r_{0}\alpha\right)^{2}
}.
\end{equation}
As we already know from \ref{conditionA)} for belly generation is necessary
\begin{equation}\label{inequality}
    -\frac{2A}{b^2} + \frac{\alpha}{r_0} <0
\end{equation}
It's reasonable to argue that when this equation is equal to zero, the space-time is in the imminence to a double-throat, then for the belly region to be formed we need a density greater than
\begin{equation}
    \rho_c = \frac{1}{8\pi(A+r_0\alpha)^2}.
\end{equation}
This result indicates that specific relations among the parameters exist for which Eq.~\eqref{rhocentral} becomes positive and sufficiently large to sustain the belly region. In this regime, the effective matter content behaves similarly to dark energy, since, as shown below, the radial pressure at the center is negative. This negative pressure generates a repulsive effect that favors the formation and stabilization of the belly. In Fig.~\ref{parameterspace2}, we display the parameter space $(A,\alpha)$, where the hatched region denotes the parameter values for which the central density is positive, together with the constraint imposed by Eq.~(\ref{conditionA)}).

The radial pressure is given by
\begin{align}
p_r(\ell) &= 
\frac{
-1 + \ell^2\left[
        -\dfrac{2 A b^{2}}{(b^{2}+\ell^{2})^{2}}
        + \dfrac{\alpha}{\left(\ell^{2}+r_{0}^{2}\right)^{1/2}
     }\right]^{2}
}{
8\pi \left[
      \dfrac{A b^{2}}{b^{2}+\ell^{2}}
      + \alpha\left(\ell^{2}+r_{0}^{2}\right)^{1/2}
\right]^{2}
},
\label{eq:pr_full}
\end{align}
At the center of the configuration, the radial pressure behaves as
\begin{equation}
p_r(\ell \to 0)\approx -\frac{1}{8\pi (A+r_0\alpha)^2}.
\end{equation}
This result shows that the pressure at the core is negative and ultimately provides the mechanism responsible for the inflation of the interthroat region, \emph{i.e.}, the formation of the belly. The corresponding equation-of-state parameter at the center, defined as $\omega_r(0) = p_r(0)/\rho(0)$, is given by
\begin{equation}
\omega_r(0) = -\frac{1}{ \left[1 - 2\left(-\frac{2A}{b^2} + \frac{\alpha}{r_0}\right)(A + r_0 \alpha)\right]}
\end{equation}

which remains negative provided that the condition~\eqref{inequality} is satisfied.

The tangential pressure is given by
\begin{align}
p_t(\ell) &= 
\frac{
-\dfrac{2A b^2 \left(b^{2}-3\ell^{2}\right)}{(b^{2}+\ell^{2})^{3}}
+ \dfrac{\alpha\,\,r_{0}^{2}}{\left(\ell^{2}+r_{0}^{2}\right)^{3/2}} 
}{
8\pi \left[
      \dfrac{A b^{2}}{b^{2}+\ell^{2}}
      + \alpha\left(\ell^{2}+r_{0}^{2}\right)^{1/2}
\right]
}.
\label{eq:pt_full}
\end{align}

A property of the present solution emerges when analyzing the global behavior of the fluid equation of state. By performing a spatial average of the equation of state parameters {$\omega_i(\ell) = {p_i}/{\rho}$} along the proper radial distance $\ell$, defined as 
\begin{equation}
\langle \omega_i \rangle = \lim_{L \to \infty} \frac{1}{L} \int_0^L \omega_i(\ell) d\ell,
\end{equation}
we find that the values converge exactly to $\langle \omega_r \rangle = -1$ and $\langle \omega_t \rangle = 0$. This result confirms that the deviations from the string cloud equation of state are spatially confined to the throat region. As the distance increases, the integrated contribution of the core fluctuations becomes negligible compared to the asymptotic background. Consequently, the object is identified globally as a structure threaded by a string cloud.
\subsection{Wormhole source}
The action governing our system is \cite{Bronnikov1973,Ellis1973}
\begin{equation}
     S[g_{\mu\nu},\phi] = \int\sqrt{-g}(R-2K\partial_{\mu}\phi\partial^{\mu}\phi)
\end{equation}
Therefore, minimizing the above action, we get the respective scalar field dynamics in curved spaces
\begin{equation}
    \partial_{\mu}(\sqrt{-g}\,g^{\mu\nu}\partial_{\nu}\phi)=0
\end{equation}
Considering that the scalar field which sustains the wormholes is function of \(\ell\), \textit{i.e}, \(\phi=\phi(\ell)\), then the field equations is
\begin{equation}
    \phi' = \frac{c}{r(\ell)^2}
\end{equation}
Where c is a constant. If we remove the disturbance, we recover the result calculated in \cite{Crispim2025}. Within the disturbance situation, the scalar field equation is
\begin{equation}
    \phi = \phi_0+c\int_{0}^{\ell}\frac{d\ell}{\left(\alpha\sqrt{\ell^2+r_0^2}+\frac{A}{1+\frac{\ell^2}{b^2}}\right)^2}
\end{equation}

\begin{figure}[h]
    \centering
    \includegraphics[width=0.45\linewidth]{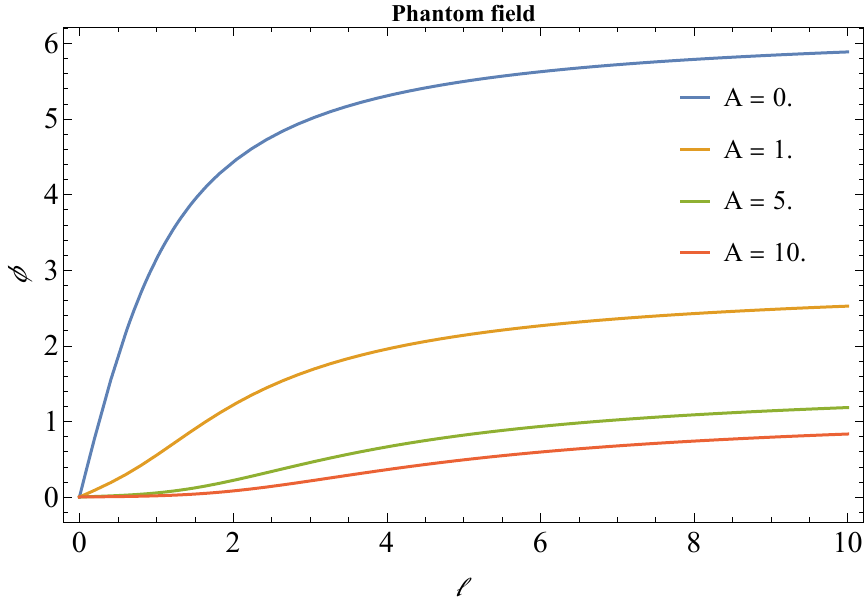}
 \includegraphics[width=0.45\linewidth]{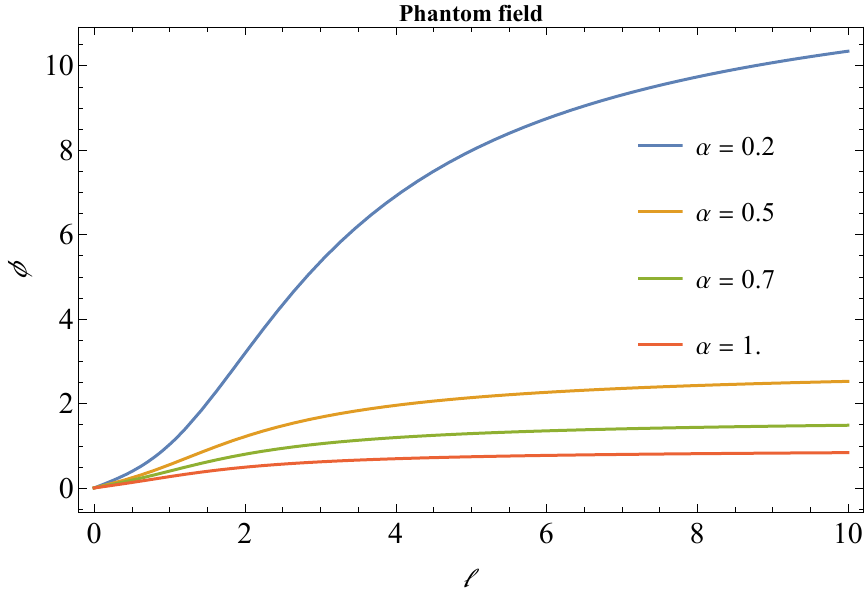}
 
    \caption{Comparative analysis of phantom field varying for differents \(\alpha\) and A. The left panel display the phantom field as functions of the proper distance $\ell$ for different perturbation amplitudes $A$, for $r_0 = 1, b=1,\alpha=0.5$, and $\phi_0 = 0$. The right panel show the behavior of the phantom field for various string cloud parameters $\alpha$ with a fixed amplitude, $A=1$. }
    \label{fig: Phantomfield}
\end{figure}

Analyzing the figure \ref{fig: Phantomfield}, we can see that as we intensify the perturbation amplitude, the phantom field tends to decrease; this can be interpreted as the perturbation decreasing the need of the phantom field to sustain the wormhole. Differently, as we increase the angular deficit, that is, decrease the \(\alpha\), we conclude that the phantom field supporting the wormhole need to be bigger.  

\subsection{Energy Conditions}

A central issue in wormhole physics is the violation of energy conditions required to maintain the throat transversability. In the context of General Relativity, the flare-out condition typically demands the violation of the Null Energy Condition (NEC), defined by $T_{\mu\nu}k^\mu k^\nu \ge 0$ for any null vector $k^\mu$. For the diagonal stress-energy tensor associated with the metric \eqref{eq:metric}, this condition implies two simultaneous inequalities:
\begin{align}
    \text{NEC}_1 &: \quad \rho + p_r \ge 0 \,, \label{eq:NEC1} \\
    \text{NEC}_2 &: \quad \rho + p_t \ge 0 \,. \label{eq:NEC2}
\end{align}
Substituting the field equations derived in Section \ref{field_eqs}, we obtain compact analytical expressions for these sums in terms of the profile function $r(\ell)$:
\begin{align}
    8\pi(\rho + p_r) &= -\frac{2r''(\ell)}{r(\ell)} \,, \label{eq:NEC_radial_analytic} \\
    8\pi(\rho + p_t) &= \frac{1 - r'(\ell)^2 - r(\ell)r''(\ell)}{r(\ell)^2} \,. \label{eq:NEC_tangential_analytic}
\end{align}
Equation \eqref{eq:NEC_radial_analytic} reveals a direct link between the radial NEC and the extrinsic curvature of the embedding surface. The condition $\rho + p_r < 0$ is strictly required only in regions where the geometry is concave upward ($r''(\ell) > 0$), i.e., at the throats.
\begin{figure}[h]
    \centering
    \includegraphics[width=0.45\linewidth]{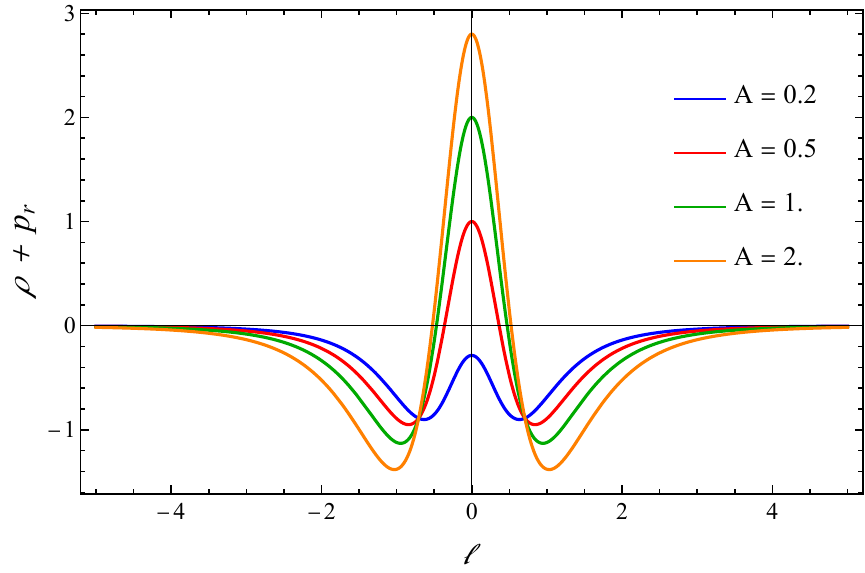}
 \includegraphics[width=0.45\linewidth]{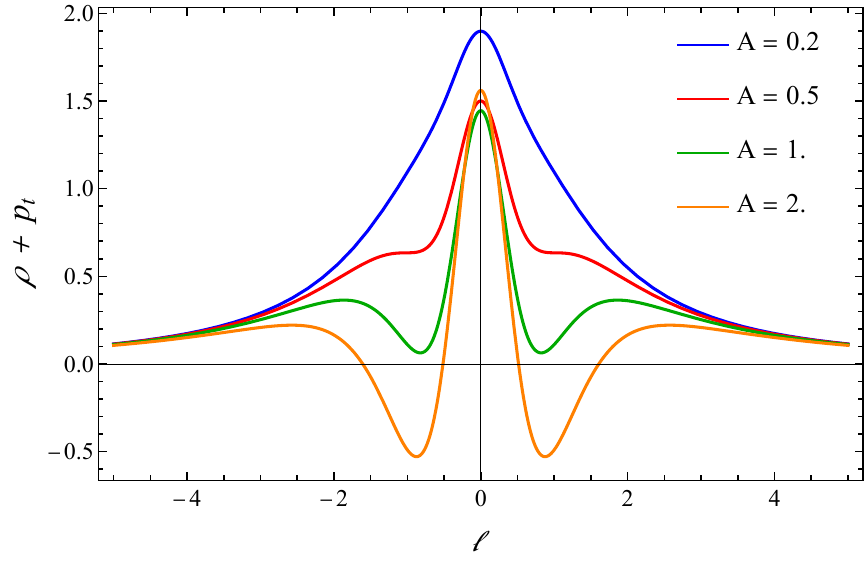}
  \includegraphics[width=0.45\linewidth]{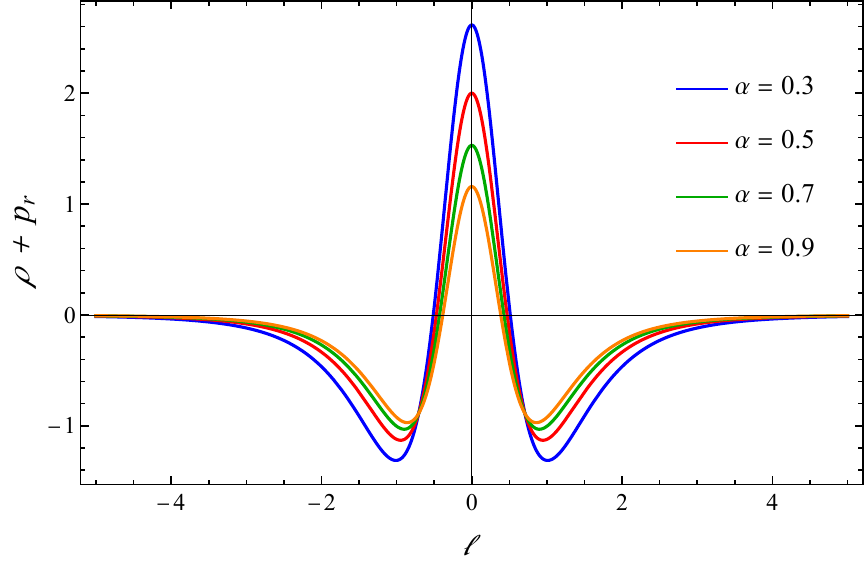}
   \includegraphics[width=0.45\linewidth]{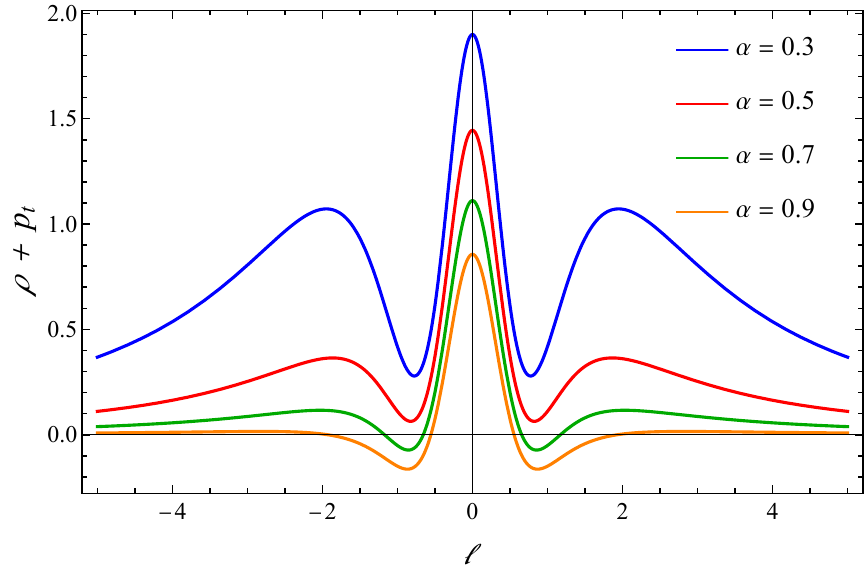}
    \caption{Comparative analysis of null energy conditions. The top panels display the NEC radial $\rho+p_r$ (left) and the NEC tangential $\rho+p_t$ (right) as functions of the proper distance $\ell$ for different perturbation amplitudes $A$, for $r_0 = 1, b=1,\alpha=0.5$. The bottom panels show the behavior of NEC radial $\rho+p_r$ (left) and NEC tangential $\rho+p_t$ (right) for various string cloud parameters $\alpha$ with a fixed amplitude, $A=1$.   }
    \label{fig: Energy_conditions}
\end{figure}

 As illustrated in the Fig. \ref{fig: Energy_conditions}, in the double-throat regime, both NEC conditions were satisfied in the belly regions' core and the exoticity is concentrated in the region near the throats. Also, we can note an interesting relationship between A, $\alpha$ and NEC matching, If A becomes very large, the tangential NEC is violated, whereas the region where the radial NEC holds expands. The same reasoning applies to $\alpha$.
\section*{CONCLUSIONS}
In this work, we have constructed a novel class of static, spherically symmetric traversable wormholes exhibiting a double-throat topology embedded in a string-cloud-like background. The model was obtained as a localized deformation of the Ellis–Bronnikov geometry under the zero-tidal-force condition, allowing a fully analytical determination of the matter content.

We have shown that the parameters controlling the background geometry plays a decisive role in the geometric transition from a single throat to a double-throat configuration. This process is orchestrated by a positive energy density in the belly core that acts to expand the space-time instead of contracting it, in the same way as dark energy, since in this case the radial pressure is negative. For higher-order generalized geometries, the double-throat emerges naturally from arbitrarily small localized perturbations, indicating that complex internal structures are not fine-tuned but rather geometrically favored.

Two key results are the reduced need for the phantom field due to the perturbation and the geometric concentration of exotic matter. As the perturbation increase, the range of \(\ell\) values where the scalar is near to zero increase, meaning that the perturbation stabilizes that region and decrease the phantom field need. The NEC violation required for the maintenance of the throat is restricted to narrow regions surrounding the two throats, while the central core satisfies all classical energy conditions, but care must be taken with large values of $\alpha$ and A, because as $\alpha$ increases, the need for the phantom field increases and the tangential condition presented a violation when these quantities increase too much. Even with this detail, this behavior is very positive and contrasts with many single-throat wormhole models, where exoticity extends over larger domains.

Asymptotically, the spacetime approaches that of a string cloud, characterized by an energy density decaying as $r^{-2}$ and an effective equation of state compatible with radially oriented string matter. This provides a physically interpretable far-field source rather than an ad hoc exotic fluid.

Although the spacetime is not asymptotically flat due to the presence of the string cloud, which results in a global monopole, this feature should be regarded as an intrinsic aspect of the physical setup rather than a limitation. The wormhole is naturally embedded in a nontrivial topological background, which plays a crucial role in sustaining the interthroat region with ordinary matter and in confining the null energy condition violation to the vicinity of the throats.

Overall, the model shows that multi-throat wormholes can arise naturally within General Relativity and that their internal topology offers a powerful mechanism for reducing and localizing exotic matter. These findings open new directions for studying the observational signatures, perturbative stability, and possible formation mechanisms of complex wormhole geometries.

\section*{Acknowledgments}
CRM would like to thank Conselho Nacional de Desenvolvimento Cient\'{i}fico e Tecnol\'ogico (CNPq) for the partial financial support, through grant 301122/2025-3. MSC and YA thank to Fundação Cearense de Apoio ao Desenvolvimento Científico e Tecnológico (FUNCAP) through the contract number ICA-0252-00030.01.68/25.

\bibliographystyle{apsrev4-2}
\bibliography{ref}

\end{document}